\documentclass[conference,letterpaper]{IEEEtran}

\usepackage{amsmath,amsfonts,amssymb,amsthm}
\usepackage{bm}
\usepackage{tikz}
\usepackage{circuitikz}
\usepackage{graphicx}
\usepackage{cite}
\usepackage{xcolor}

\title{Stacked Intelligent Metasurfaces (SIM) in the Nonlinear Regime: A Multiport Network Model Approach}
\author{Andrea Abrardo, Alberto Toccafondi%
\thanks{Andrea Abrardo and Alberto Toccafondi are with the Department of Information Engineering and Mathematics (DIISM), University of Siena, 53100 Siena, Italy, and with CNIT, Consorzio Nazionale Interuniversitario per le Telecomunicazioni (e-mail: abrardo@unisi.it; alberto.toccafondi@unisi.it).}}
\date{}

\begin{document}
\maketitle
\pagestyle{empty}

\title{Stacked Intelligent Metasurfaces (SIM) in the Nonlinear Regime: A Multiport Network Model Approach}
\author{Andrea Abrardo, Alberto Toccafondi%
\thanks{Andrea Abrardo and Alberto Toccafondi are with the Department of Information Engineering and Mathematics (DIISM), University of Siena, 53100 Siena, Italy, and with CNIT, Consorzio Nazionale Interuniversitario per le Telecomunicazioni (e-mail: abrardo@unisi.it; alberto.toccafondi@unisi.it).}}
\date{}

\maketitle

\begin{abstract}
We present a physically consistent multiport framework for stacked intelligent
metasurfaces (SIMs) with linear and explicit nonlinear terminations. The model
provides closed-form input--output relations in the linear case and fixed-point
forward evaluation in the nonlinear case, with adjoint-based gradients for
optimization in both settings. Under stage-isolated SIM structure, complexity
remains $\mathcal{O}(QK^3)$. In a 28 GHz near-field localization case study,
nonlinear terminations improve transfer-function matching and reduce mean localization error,
close to the ideal benchmark.
\end{abstract}

\begin{IEEEkeywords}
Stacked intelligent metasurfaces, multiport network model, nonlinear
terminations, adjoint optimization, near-field localization.
\end{IEEEkeywords}

%=================================================
\section{Introduction}

\subsection{Background and Motivation}

Stacked Intelligent Metasurfaces (SIMs) are emerging as a key physical-layer
architecture for wave-domain processing in beyond-5G/6G systems. By stacking
multiple programmable transmissive layers and jointly tuning their responses,
SIMs offer a richer electromagnetic control space than conventional single-layer
RIS platforms, with applications to communication, sensing, and localization
~\cite{DiRenzo:19b,DiRenzo2020_JSAC,RenzoDT22,10158690,10922857,An_2024_ref4}.

The SIM concept has rapidly evolved from system-level intuition to more
structured design methodologies. Early models often adopted convenient cascaded
channel abstractions. While useful for tractable optimization, such models may
not fully capture inter-port coupling, internal reflections, and boundary
conditions when dense electromagnetic interactions are relevant. This has
motivated physically consistent network formulations based on scattering
parameters and multiport theory, which provide a tighter link between EM
behavior and end-to-end signal processing~\cite{DR1,DR2,abrardo_E,Abrardo24RisOpt}.

Within this line of research, only a limited number of works currently address
SIM architectures through full multiport modeling. A key application scenario
is SIM-aided near-field localization under a multiport-network formulation,
as reported in ~\cite{abrardo_Eusipco}.

In parallel, recent nonlinear EM-based signal processing results have shown the
significant potential of nonlinear hardware in SIM-aided architectures
~\cite{Fabiani2025NonlinearEM}. This evidence motivates extending physically
consistent SIM models beyond linear terminations, so as to combine nonlinear
wave manipulation with rigorous electromagnetic network descriptions.

\subsection{Paper Contributions}

Building on this context, this paper develops a modernized SIM multiport
framework with explicit input--output and adjoint-gradient expressions under
structured diagonal T-RIS terminations, and extends it to nonlinear cell
behaviors. The proposed methodology differs from simplified cascade-only SIM
models by preserving a complete Tx--SIM--Rx multiport representation with
explicit nonlinear termination laws.

The contributions can be summarized as follows. First, we formulate the
nonlinear SIM internal interaction as a fixed-point problem, which yields a
computationally tractable forward input--output evaluation while retaining
physical consistency at the network level. Second, we derive an adjoint-based
gradient computation tailored to the nonlinear multiport setting, enabling
scalable parameter optimization for large SIM configurations. Third, we provide
a unified analytical pipeline that bridges physically consistent SIM network
modeling and nonlinear wave-domain signal processing, thus extending the design
space highlighted by recent nonlinear SIM studies to a full multiport
Tx--SIM--Rx treatment.

As a consequence, the proposed framework offers both modeling fidelity and
algorithmic practicality: it captures internal propagation and inter-port
interactions within the SIM, and at the same time supports gradient-driven
optimization workflows for communication/sensing-oriented objectives.

%=================================================
\section{S-parameter linear SIM model, transfer function, and gradient evaluation}
\label{sec:SIM_linear_multiport}

\begin{figure}[t]
  \centering
   \includegraphics[width=1\columnwidth]{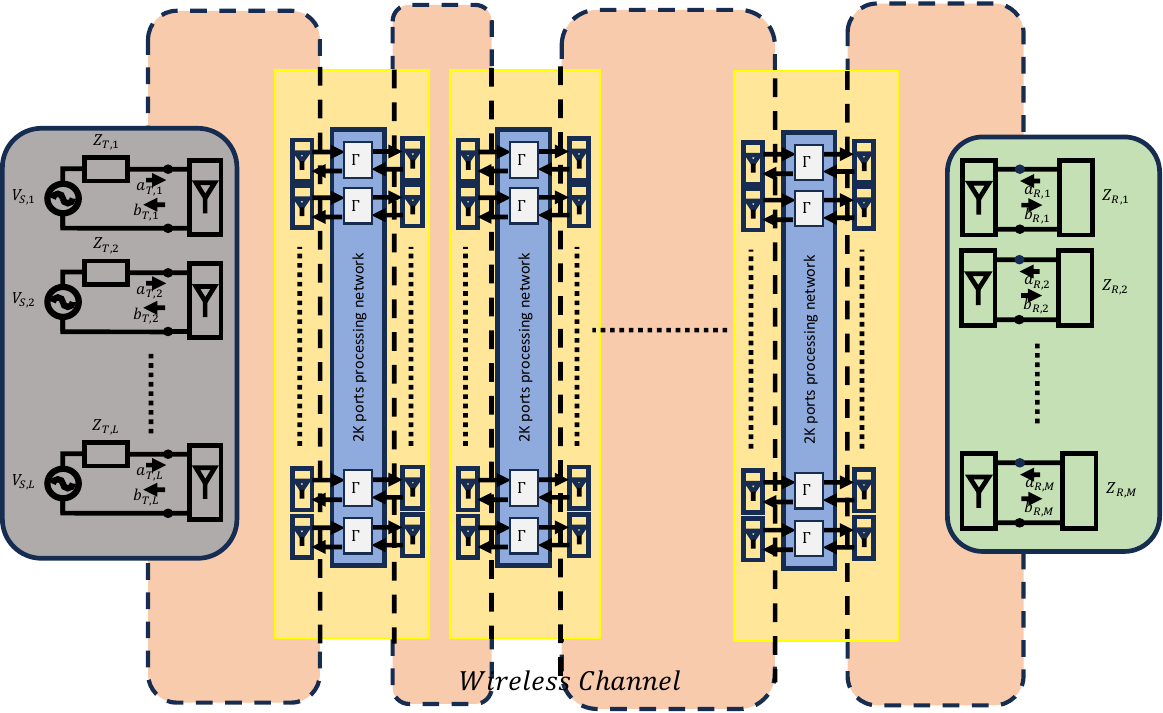}
  \caption{Illustration of a SIM architecture as a multiport network. The wireless channel couples the transmitter to the first SIM stage, then couples facing layers within adjacent T-RIS stages, and finally couples the last SIM stage to the receiver, while electromagnetic interactions between non-adjacent T-RIS stages are neglected.}
  \label{fig:simI_schematic}
\end{figure}

We consider the stacked intelligent metasurface (SIM) architecture in
Fig.~\ref{fig:simI_schematic}, composed of $Q$ transmissive reconfigurable
intelligent surfaces (T-RISs) arranged in a layered (stacked) structure.
Each stage is a T-RIS made of two facing layers with $K$ ports per layer.
The SIM therefore comprises $N=2QK$ internal ports, and is embedded between a
transmitting array (Tx) with $L$ ports and a receiving array (Rx) with $M$ ports.
The overall Tx--SIM--Rx system is modeled as a linear $N_t=L+N+M$-port network
described by its scattering matrix $\bm S\in\mathbb{C}^{N_t\times N_t}$:
\begin{equation}
\bm b = \bm S \bm a,
\qquad \bm a,\bm b\in\mathbb{C}^{N_t},
\label{eq:SIM_S_global}
\end{equation}
where $\bm a$ and $\bm b$ denote the incident and reflected power-wave vectors,
respectively.

\subsection{Port partitioning and boundary conditions}
We partition the ports into three groups:
\begin{align}
\bm a =&
\begin{bmatrix}\bm a_T \\ \bm a_E \\ \bm a_R\end{bmatrix},
\qquad
\bm b =
\begin{bmatrix}\bm b_T \\ \bm b_E \\ \bm b_R\end{bmatrix},
\qquad \\
\bm S=&
\begin{bmatrix}
\bm S_{TT} & \bm S_{TE} & \bm S_{TR}\\
\bm S_{ET} & \bm S_{EE} & \bm S_{ER}\\
\bm S_{RT} & \bm S_{RE} & \bm S_{RR}
\end{bmatrix}
\end{align}
As in \cite{abrardo_E}, we assume {the ports of the transmitter and the receiver matched for zero reflection}:
\[
\bm a_T=\bm a_S,\qquad \bm a_R=\bm 0.
\]
The internal and output relations then read
\begin{align}
\bm b_E &= \bm S_{ET}\bm a_S + \bm S_{EE}\bm a_E,
\label{eq:bE_general}\\
\bm y \triangleq \bm b_R &= \bm S_{RT}\bm a_S + \bm S_{RE}\bm a_E.
\label{eq:y_general_vec}
\end{align}

\subsection{SIM terminations: diagonal T-RIS and structure of $\bm\Gamma(\boldsymbol\eta)$}
The SIM internal ports are ``closed'' by the programmable T-RIS circuitry.
In the most general form, this is modeled as an $N$-port termination network
\begin{equation}
\bm a_E = \bm \Gamma(\boldsymbol\eta)\,\bm b_E,
\label{eq:SIM_Gamma_global}
\end{equation}
parametrized by a control vector $\boldsymbol\eta$.

We focus on the widely adopted \emph{diagonal T-RIS} architecture
\cite{Abra_SIM1}, where the SIM comprises $QK$ independent
tunable two-port cells. Index the cells by $p\in\{1,\ldots,QK\}$ and denote by
$m(p)$ and $n(p)$ the two internal ports coupled by the $p$-th cell (located on
opposite sides of the same stage). The cell relation reads
\begin{equation}
\begin{bmatrix}
a_{E,m(p)}\\[2pt]
a_{E,n(p)}
\end{bmatrix}
=
\bm\Gamma_{\mathrm{cell}}(\eta_p)
\begin{bmatrix}
b_{E,m(p)}\\[2pt]
b_{E,n(p)}
\end{bmatrix},
\label{eq:SIM_cell}
\end{equation}
where $\bm\Gamma_{\mathrm{cell}}(\eta_p)\in\mathbb{C}^{2\times 2}$ models a
generally non-ideal tunable two-port. The ideal phase-shifter model is recovered
as
\begin{equation}
\bm\Gamma_{\mathrm{cell}}(\eta_p)=
\begin{bmatrix}
0 & e^{j\eta_p}\\
e^{j\eta_p} & 0
\end{bmatrix}.
\label{eq:SIM_ideal_cell}
\end{equation}

Collecting all cells, $\bm\Gamma(\boldsymbol\eta)\in\mathbb{C}^{N\times N}$ is
sparse and block-diagonal at the cell level (non-overlapping $2\times 2$ blocks),
consistent with the layered diagonal T-RIS structure in Fig.~\ref{fig:simI_schematic}.
Its inverse $\bm\Gamma^{-1}(\boldsymbol\eta)$ preserves the same sparsity and is
obtained by inverting each $2\times2$ cell block independently (when invertible).

\subsection{End-to-end transfer function}
With straightforward algebraic manipulations, one can
verify that the resulting end-to-end input--output relation can be written as
\begin{subequations}\label{eq:GTF_def_with_direct}
\begin{align}
\bm y = \big(\bm S_{RT} +\bm S_{RE}\,\bm T(\boldsymbol{\eta})\,\bm S_{ET}\big)\bm a_S,\\
\bm T(\boldsymbol{\eta}) = (\bm I-\bm\Gamma(\boldsymbol{\eta})\bm S_{EE})^{-1}\bm\Gamma(\boldsymbol{\eta}).
\end{align}
\end{subequations}

\subsection{Loss function}
For $I$ excitations collected in $\bm A_S=[\bm a_S^{(1)},\ldots,\bm a_S^{(I)}]$,
let $\bm Y=[\bm y^{(1)},\ldots,\bm y^{(I)}]$ denote the corresponding outputs and
$\bm Y_d$ the desired targets. With a complex scaling $\beta\in\mathbb{C}$,
define $\bm E\triangleq \beta\bm Y-\bm Y_d$ and the loss
\begin{equation}
L=\|\bm E\|_F^2,
\qquad
\langle \bm X,\bm Y\rangle \triangleq \mathrm{trace}(\bm X^H\bm Y).
\label{eq:SIM_loss}
\end{equation}

\subsection{Gradient evaluation}
The gradient evaluation follows the standard adjoint-state structure. The forward fields satisfy
\begin{equation}
(\bm I-\bm S_{EE}\bm\Gamma)\bm B_E = \bm S_{ET}\bm A_S,
\label{eq:SIM_forward_BE}
\end{equation}
and the differential relation
\begin{equation}
(\bm I-\bm\Gamma\bm S_{EE})
\frac{\partial\bm A_E}{\partial\eta_p}
=
\frac{\partial\bm\Gamma}{\partial\eta_p}\bm B_E.
\label{eq:SIM_dAE}
\end{equation}
Using $\delta\bm Y=\bm S_{RE}\delta\bm A_E$, the first-order loss variation is
\begin{equation}
\delta L
=
2\,\Re\!\left\langle
\bm S_{RE}^H(\beta^*\bm E),
\delta\bm A_E
\right\rangle.
\label{eq:SIM_deltaL}
\end{equation}
Define the adjoint forcing
\begin{equation}
\bm q\triangleq \bm S_{RE}^H(\beta^*\bm E),
\label{eq:SIM_q}
\end{equation}
and the adjoint variable $\bm U\in\mathbb{C}^{N\times I}$ as the solution of
\begin{equation}
(\bm I-\bm\Gamma\bm S_{EE})^H\bm U=\bm q.
\label{eq:SIM_adj}
\end{equation}
Then, the gradient is
\begin{equation}
\frac{\partial L}{\partial\eta_p}
=
2\,\Re\!\left\langle
\bm U,\,
\frac{\partial\bm\Gamma}{\partial\eta_p}\bm B_E
\right\rangle.
\label{eq:SIM_grad}
\end{equation}

As shown in~\cite{Abra_SIM1}, the complexity of the gradient computation,
and therefore of the overall optimization algorithm, scales as
\begin{equation}
\mathcal{C}_{\mathrm{SIM}}=\mathcal{O}(QK^3),
\end{equation}
i.e., cubic in the number of ports per layer and linear in the number of
stacked stages.

%=================================================
%=================================================
\section{Nonlinear explicit SIM termination: I/O evaluation and gradient computation}
\label{sec:SIM_nonlinear_explicit}

This section extends the linear S-parameter SIM model in
Section~\ref{sec:SIM_linear_multiport} to the case in which each diagonal
T-RIS cell is described by an \emph{explicit nonlinear} termination law.
The goal is twofold: i) compute the end-to-end input--output map
$\bm y=\bm b_R$ for given excitations (forward evaluation), and ii) compute the
gradient of a quadratic loss with respect to the SIM control parameters
$\boldsymbol\eta$, by an adjoint method.

\subsection{Explicit nonlinear termination law}
\label{subsec:explicit_nonlin_law}

In the linear case, the programmable SIM termination is
$\bm a_E=\bm\Gamma(\boldsymbol\eta)\bm b_E$ and yields the closed-form transfer
function \eqref{eq:GTF_def_with_direct}. Here we assume instead an
\emph{explicit nonlinear} map
\begin{equation}
\bm a_E = \bm f\!\big(\bm b_E;\bm\Gamma(\boldsymbol\eta)\big),
\label{eq:explicit_map_global}
\end{equation}
where $\bm f:\mathbb{C}^N\to\mathbb{C}^N$ may depend on the (sparse) termination
matrix $\bm\Gamma(\boldsymbol\eta)$ and typically inherits the diagonal T-RIS
cell separability.

The adoption of an \emph{explicit} relation of the form
$\bm a_E=\bm f(\bm b_E;\bm\Gamma)$ implicitly assumes that each tunable cell
can be modeled as a two-port network that is locally adapted at its ports,
in analogy with the ideal phase-shifter model
\begin{equation}
\bm\Gamma_{\mathrm{cell}}(\eta_p)=
\begin{bmatrix}
0 & e^{j\eta_p}\\
e^{j\eta_p} & 0
\end{bmatrix},
\end{equation}
for which $S_{11}=S_{22}=0$ and
\begin{equation}
a_{E,m(p)} = e^{j\eta_p}\, b_{E,n(p)}, 
\qquad
a_{E,n(p)} = e^{j\eta_p}\, b_{E,m(p)}.
\end{equation}
In this ideal case, no internal reflections occur at the cell ports and the
termination law is purely transmissive.

The explicit nonlinear law \eqref{eq:explicit_map_global} can be interpreted as
a nonlinear generalization of this matched two-port behavior, where the
transmitted wave may depend nonlinearly on the magnitude (or other features)
of the incident wave. This modeling choice will be made more concrete in the
specific example treated in the sequel. Nevertheless, it does not reduce the
generality of the framework, since any physically realizable two-port can, in
principle, be embedded within an adapted network representation, allowing its
effective behavior to be described by an explicit relation between incident and
reflected waves at the SIM internal ports.

The forward equation \eqref{eq:bE_general} combined with
\eqref{eq:explicit_map_global} yields the reduced fixed-point system
\begin{equation}
\bm a_E
=
\bm f\!\big(\bm S_{ET}\bm a_S+\bm S_{EE}\bm a_E;\bm\Gamma(\boldsymbol\eta)\big).
\label{eq:fixed_point_AE}
\end{equation}

\subsection{Forward evaluation: iterative I/O computation}
\label{subsec:forward_iter_IO}

For each excitation $\bm a_S$, the output is
\begin{equation}
\bm y(\bm a_S,\boldsymbol\eta)
=
\bm S_{RT}\bm a_S + \bm S_{RE}\,\bm a_E(\bm a_S,\boldsymbol\eta),
\label{eq:IO_nl_def}
\end{equation}
where $\bm a_E(\bm a_S,\boldsymbol\eta)$ is obtained by solving
\eqref{eq:fixed_point_AE}. Since \eqref{eq:fixed_point_AE} is nonlinear, $\bm a_E$
is computed numerically via a fixed-point iteration (optionally relaxed):
\begin{align}
\bm b_E^{(t)} &= \bm S_{ET}\bm a_S + \bm S_{EE}\bm a_E^{(t)},
\label{eq:fp_bE}\\
\bm a_{E,\mathrm{map}}^{(t)} &= \bm f\!\big(\bm b_E^{(t)};\bm\Gamma(\boldsymbol\eta)\big),
\label{eq:fp_map}\\
\bm a_E^{(t+1)} &= (1-\omega)\bm a_E^{(t)}+\omega\,\bm a_{E,\mathrm{map}}^{(t)},
\qquad \omega\in(0,1].
\label{eq:fp_relax}
\end{align}
The iteration stops when the residual
$\|\bm a_E^{(t)}-\bm a_{E,\mathrm{map}}^{(t)}\|_2$ falls below a tolerance.

The dominant cost per iteration is the evaluation of the linear coupling term
\eqref{eq:fp_bE}, i.e., the matrix--vector (or matrix--matrix, for multiple
excitations) product $\bm S_{EE}\bm a_E^{(t)}$. Owing to the layered SIM
architecture, $\bm S_{EE}$ is not dense: under the stage-isolated model it has a
block-banded structure, where each stage contributes a local block (of size
$K\times K$ for edge stages and $2K\times 2K$ for intermediate stages) and
coupling occurs only between adjacent layers.

As a consequence, each row of $\bm S_{EE}$ contains nonzero entries only within
a limited set of columns associated with the corresponding stage (and possibly
its immediate neighbor). In particular, there are at most $2K$ nonzero columns
per row. Therefore, the product $\bm S_{EE}\bm a_E^{(t)}$ requires at most
$\mathcal{O}(2NK)$ operations, rather than $\mathcal{O}(N^2)$ as in the dense
case.

Since the nonlinear map $\bm f(\cdot)$ acts componentwise (or at most on
two-port cell pairs in the diagonal T-RIS structure), its evaluation scales
linearly with $N$. Hence, the overall per-iteration complexity of the fixed-point
solver is dominated by the structured multiplication
$\bm S_{EE}\bm a_E^{(t)}$ and scales as
\[
\mathcal{O}(2NK),
\]
which is linear in the total number of SIM ports $N$ for fixed $K$.

\paragraph{Multiple excitations.}
Given $I$ excitations collected in $\bm A_S=[\bm a_S^{(1)},\ldots,\bm a_S^{(I)}]$,
the same iteration applies column-wise, yielding matrices
$\bm A_E, \bm B_E$ and outputs
\begin{equation}
\bm Y = \bm S_{RT}\bm A_S + \bm S_{RE}\bm A_E.
\label{eq:Y_nl_mat}
\end{equation}

\paragraph{Remark (well-posedness).}
Nonlinear explicit terminations may admit multiple fixed points or none, and the
iteration may converge only for suitable initializations and relaxation
$\omega$. In physically meaningful regimes (e.g., passive/dissipative nonlinear
cells operating within their stable region), one typically expects a unique
steady-state solution for each admissible excitation; if the iteration converges,
the resulting fixed point is interpreted as the physically realized response.

\subsection{Adjoint gradient for explicit nonlinear terminations}
\label{subsec:adjoint_grad_explicit}

In the linear case the SIM termination is
\begin{equation}
\bm a_E=\bm\Gamma(\boldsymbol\eta)\bm b_E,
\label{eq:lin_gamma}
\end{equation}

and the gradient follows directly from differentiating this linear relation.

We now consider instead an explicit nonlinear cell-wise map
\begin{equation}
\bm a_E=\bm f(\bm b_E,\boldsymbol\eta),
\label{eq:explicit_local_map}
\end{equation}
where, due to the diagonal T-RIS architecture, the map acts independently on
each two-port cell $m(p)$ and $n(p)$.

\subsubsection{Linearized forward sensitivity via the chain rule}

Recall that
\[
\bm b_E = \bm S_{ET}\bm a_S + \bm S_{EE}\bm a_E.
\]

Differentiating \eqref{eq:explicit_local_map} with respect to $\eta_p$
and applying the ordinary chain rule yields
\begin{equation}
\frac{\partial \bm a_E}{\partial \eta_p}
=
\frac{\partial \bm f}{\partial \bm b_E}
\frac{\partial \bm b_E}{\partial \eta_p}
+
\frac{\partial \bm f}{\partial \eta_p}.
\label{eq:chain_rule_global}
\end{equation}

Define the Jacobian matrix
\[
\bm J_b \triangleq \frac{\partial \bm f}{\partial \bm b_E}
\in\mathbb{C}^{N\times N},
\]
evaluated at the converged forward solution. Note that, Although $\bm J_b$ is formally an $N\times N$ matrix, it is highly sparse due to
the diagonal T-RIS (cell-wise) separability of $\bm f$.
In particular, each output component $a_{E,i}$ depends only on the wave(s) at
the two ports of its own cell. Hence, the $i$-th row of $\bm J_b$ has non-zero
entries only in the columns corresponding to that local port pair.

In the matched case
$a_{E,m(p)}=f_m(b_{E,n(p)},\eta_p)$ and
$a_{E,n(p)}=f_n(b_{E,m(p)},\eta_p)$,
each row contains exactly one non-zero entry, so $\bm J_b$ has $\mathcal{O}(N)$
non-zeros (approximately $N$) and can be seen as a permuted diagonal matrix.
In the more general (possibly mismatched) two-port case where each output may
depend on both $b_{E,m(p)}$ and $b_{E,n(p)}$, each row has at most two non-zero
entries, leading to $\mathcal{O}(2N)$ non-zeros (equivalently, $2\times2$
block-sparsity at the cell level).

Since $\bm a_S$ is fixed during gradient evaluation,
\[
\frac{\partial \bm b_E}{\partial \eta_p}
=
\bm S_{EE}\frac{\partial \bm a_E}{\partial \eta_p}.
\]

Substituting into \eqref{eq:chain_rule_global} gives
\begin{equation}
\frac{\partial \bm a_E}{\partial \eta_p}
=
\bm J_b \bm S_{EE}
\frac{\partial \bm a_E}{\partial \eta_p}
+
\frac{\partial \bm f}{\partial \eta_p}.
\end{equation}

Rearranging,
\begin{equation}
(\bm I-\bm J_b\bm S_{EE})
\frac{\partial \bm a_E}{\partial \eta_p}
=
\frac{\partial \bm f}{\partial \eta_p}.
\label{eq:lin_sensitivity_final}
\end{equation}

This equation is the nonlinear analogue of the linear-SIM sensitivity equation.
The only difference is that $\bm J_b$ replaces $\bm\Gamma$.

Importantly, the term $
\frac{\partial \bm f}{\partial \eta_p}$ is nonzero only on the two entries corresponding to the cell $(m(p),n(p))$.

\subsubsection{Adjoint equation}

For multiple excitations collected in $\bm A_S$, define
\[
\bm Y=\bm S_{RT}\bm A_S+\bm S_{RE}\bm A_E,
\qquad
\bm E=\beta\bm Y-\bm Y_d.
\]

Using $\delta\bm Y=\bm S_{RE}\delta\bm A_E$, the loss variation becomes
\[
\delta L
=
2\,\Re\!\left\langle
\bm S_{RE}^H(\beta^*\bm E),
\delta\bm A_E
\right\rangle.
\]

Define the adjoint forcing
\[
\bm q \triangleq \bm S_{RE}^H(\beta^*\bm E).
\]

Introduce the adjoint variable $\bm U$ as the solution of
\begin{equation}
(\bm I-\bm J_b\bm S_{EE})^H \bm U = \bm q.
\label{eq:adjoint_final}
\end{equation}

Note that the operator appearing here is the Hermitian transpose of the
forward sensitivity operator in \eqref{eq:lin_sensitivity_final}.

\subsubsection{Gradient expression}

Multiplying \eqref{eq:lin_sensitivity_final} by $\bm U$ and using
\eqref{eq:adjoint_final}, we obtain
\begin{equation}
\frac{\partial L}{\partial \eta_p}
=
2\,\Re\!\left\langle
\bm U,\;
\frac{\partial \bm f}{\partial \eta_p}
\right\rangle.
\label{eq:gradient_clean}
\end{equation}

Thus, the gradient with respect to $\eta_p$ reduces to the inner product between
the adjoint field and the local derivative of the nonlinear two-port law.
No matrix-level directional derivatives are required; the computation is entirely
based on standard scalar chain-rule differentiation applied at the cell level.

\paragraph{Structure and computational complexity.}
The adjoint system \eqref{eq:adjoint_final}
\[
(\bm I-\bm J_b\bm S_{EE})^H \bm U = \bm q
\]
is formally identical in structure to the linear-SIM adjoint equation, the only
difference being that $\bm J_b$ replaces $\bm\Gamma$.

Under the stage-isolated SIM assumptions, the matrix $\bm S_{EE}$ is block-banded
with respect to the stage partition (each stage contributing a $2K\times 2K$
block). Moreover, as discussed above, the Jacobian $\bm J_b$ inherits the same
cell-wise sparsity structure as the diagonal T-RIS termination: each row has
non-zero entries only within its local $2\times 2$ cell block. Consequently,
$\bm J_b$ is block-diagonal at the cell level and does not introduce additional
long-range coupling across stages.

As a result, the product $\bm J_b\bm S_{EE}$ preserves the same block-banded
structure encountered in the linear case, and the operator
\[
\bm I-\bm J_b\bm S_{EE}
\]
has the same sparsity pattern as
$\bm I-\bm\Gamma\bm S_{EE}$.

Furthermore, the adjoint forcing
\[
\bm q=\bm S_{RE}^H(\beta^*\bm E)
\]
has non-zero entries only on the last $K$ ports (since $\bm S_{RE}$ is non-zero
only in its last $K$ columns). Therefore, only the last $K$ rows of
$(\bm I-\bm J_b\bm S_{EE})^{-1}$ are required for the gradient evaluation.

Exploiting this structure, the adjoint system can be solved with the same
specialized block-recursive algorithms used in the linear SIM case
(see~\cite{Abra_SIM1}), leading to an overall computational complexity
\[
\mathcal{C}_{\mathrm{SIM}}=\mathcal{O}(QK^3),
\]
i.e., cubic in the number of ports per layer and linear in the number of stages.
Hence, the introduction of an explicit cell-wise nonlinearity does not alter the
asymptotic computational scaling of the SIM optimization framework.
%========================================================

\section{Case study: anti-parallel diode limiter}
\label{subsec:amp_nl_antiparallel}

Consider a two-port cell whose dominant nonlinear element is an anti-parallel diode pair connected in shunt across the transmission line. Let $v(t)$ denote the RF voltage across the diode pair. 
In realistic diode models the current-voltage relation is implicit due to the presence of series resistance, and therefore the nonlinear current cannot be written as an explicit function of the applied voltage.

Assuming a single-tone excitation $v(t)=V\cos(\omega_0 t)$, the fundamental component of the nonlinear current is evaluated numerically by computing the diode current over one RF period and extracting the first harmonic via Fourier analysis. This procedure is equivalent to a single-tone harmonic-balance evaluation of the nonlinear shunt limiter.

Let the incident RF wave be written in complex-envelope form
\begin{equation}
b(t)=\Re\{\tilde b(t)e^{j\omega_0 t}\},
\end{equation}
with slowly varying envelope $\tilde b(t)$. Under the fundamental-only approximation and approximate matching, the limiter mainly introduces amplitude compression with negligible phase distortion. This motivates the radial envelope model
\begin{equation}
\tilde a(t)=g\!\big(|\tilde b(t)|\big)\tilde b(t),
\label{eq:radial_map}
\end{equation}
where $g(r)$ captures the amplitude-dependent attenuation.

Since the exact AM/AM characteristic cannot be expressed in closed form, the numerically evaluated response is approximated using the Rapp soft-limiter model \cite{Rapp1991}
\begin{equation}
g(r)=\frac{g_0}{\big(1+(r/r_s)^{2p}\big)^{1/(2p)}},
\qquad r\ge 0,
\label{eq:g_soft_limiter}
\end{equation}
where $g_0$ is the small-signal gain, $r_s$ is a soft saturation level, and $p$ controls the smoothness of the transition to saturation. 
The parameters $(g_0,r_s,p)$ are obtained by fitting the numerically computed AM/AM response. Although originally introduced for power amplifier nonlinearities, this model also accurately reproduces the compression behavior of the considered limiter, as verified by numerical simulations.

The model in \eqref{eq:radial_map} can be expressed directly in terms of the power-wave variables. 
Assuming approximate matching also under compression, the two-port cell with control phase $\eta_p$ is modeled as
\begin{equation}
a_{E,\substack{m(p)\\[-2pt] n(p)}} =
g\!\big(|b_{E,\substack{n(p)\\[-2pt] m(p)}}|\big)e^{j\eta_p} b_{E,\substack{n(p)\\[-2pt] m(p)}}
\end{equation}

% \begin{equation}
% a_{E,\frac{m(p)}{n(p)}} = g\!\big(|b_{E,n(p)}|\big)e^{j\eta_p} b_{E,n(p)},\~
% a_{E,n(p)} = g\!\big(|b_{E,m(p)}|\big)e^{j\eta_p} b_{E,m(p)},
% \end{equation}
which defines the nonlinear map used in the I/O mapping and gradient derivations.

%=================================================
\section{Results and Discussion}
\label{sec:results}

The reference setup follows the SIM near-field localization scenario introduced
in~\cite{abrardo_Eusipco}. The objective is to implement, through the SIM, a
polar-domain transformation that maps the incident field into resolvable
angle--range bins and then estimate the user position from the SIM output power
levels.

\begin{figure}[t]
    \centering
    \includegraphics[width=\columnwidth]{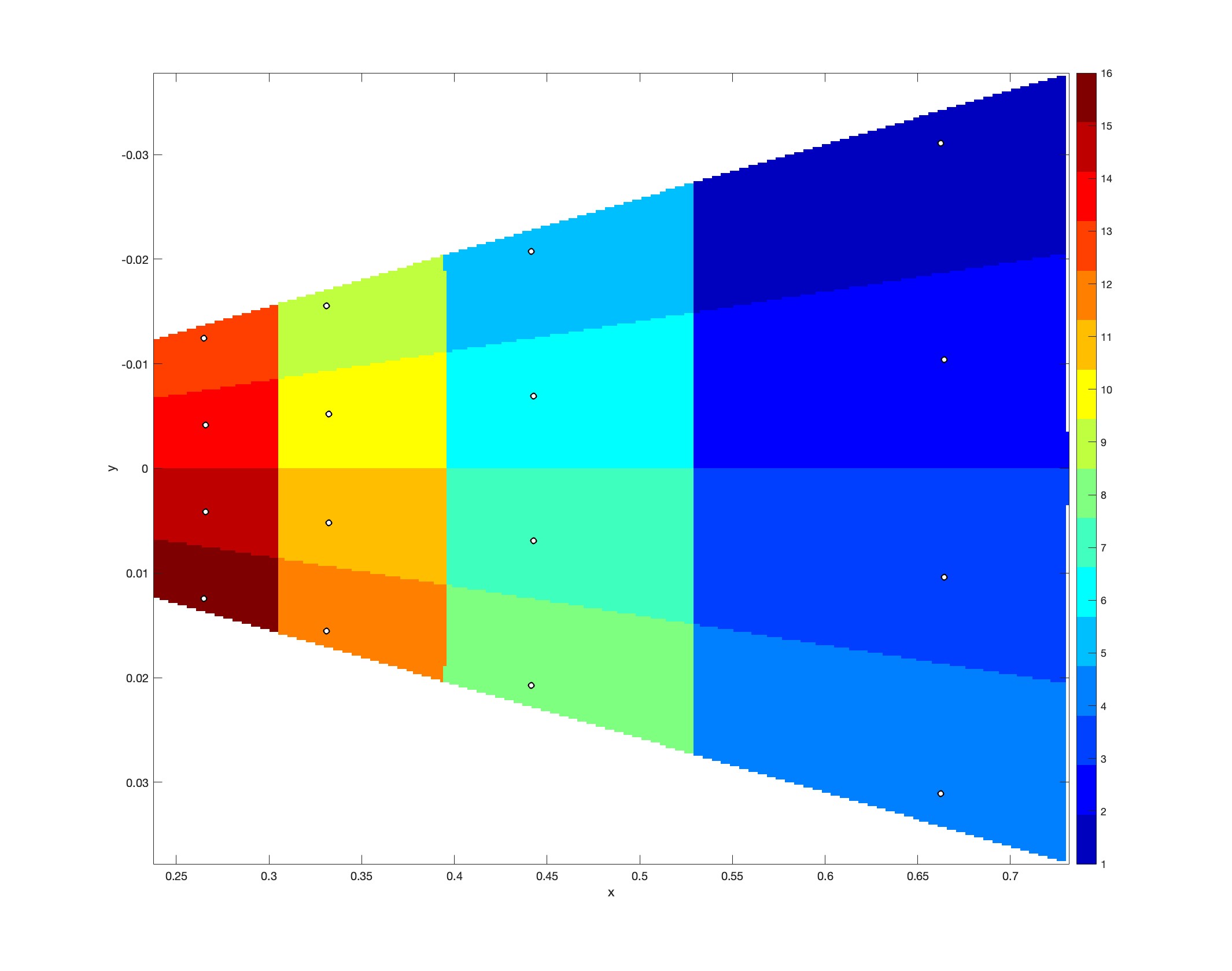}
    \caption{Top-view ideal map over the localization area of interest. Anchor points define the target transfer function, and false colors denote the dominant output channel (maximum response) at each position.}
    \label{fig:ideal_map_outputs}
\end{figure}

In particular, we consider $f_0=28$ GHz, $Q=5$ stacked layers, and a target
mapping with $4$ angle bins and $4$ range bins ($16$ output channels).
Figure~\ref{fig:ideal_map_outputs} shows the corresponding ideal top-view map
used as benchmark.

Localization is obtained from the probe powers, selecting the dominant
angle--range bin and refining the estimate by local interpolation. In
transfer-function matching, the residual error decreases from about $2\%$ (FL)
to about $1\%$ (NL). At $\mathrm{SNR}=10$ dB, the mean localization error is
$5.78$ cm (FL), $4.75$ cm (NL), and $4.17$ cm for the ideal benchmark.
These results show a consistent advantage of nonlinear terminations while
keeping the same localization pipeline.

\section{Conclusions}
This paper introduces a unified SIM framework that extends physically
consistent linear multiport modeling to explicit nonlinear cell terminations,
combining fixed-point forward evaluation and adjoint-based gradient
optimization.

The nonlinear extension preserves the same asymptotic complexity scaling as the
linear case, $\mathcal{O}(QK^3)$, while improving modeling flexibility.
Numerical results for near-field localization show consistent gains of
nonlinear SIMs over fully linear terminations in both transfer-function
matching and localization accuracy.

\bibliographystyle{IEEEtran}
\bibliography{biblio_SIM}
\end{document}